\documentclass[%
 reprint,
superscriptaddress,
 amsmath,amssymb,
 aps,
]{revtex4-1}

\usepackage{graphicx}
\usepackage{dcolumn}
\usepackage{braket}
\usepackage{gensymb}

\usepackage[symbol*]{footmisc}
\usepackage{bm}

\usepackage{color}

\DefineFNsymbolsTM{otherfnsymbols}{%
  = =
}%

\setfnsymbol{otherfnsymbols}

\begin{document}

\preprint{APS/123-QED}

\title{Protocol for Fermionic Positive-Operator-Valued Measures}
\author{D. R. M. Arvidsson-Shukur}
\thanks{These authors contributed equally to this paper.}
\affiliation{%
 Cavendish Laboratory, Department of Physics, University of Cambridge, Cambridge CB3 0HE, United Kingdom
}
\affiliation{%
Hitachi Cambridge Laboratory, J. J. Thomson Avenue, CB3 0HE, Cambridge, United Kingdom
}
\author{H. V. Lepage}
\thanks{These authors contributed equally to this paper.}
\affiliation{%
 Cavendish Laboratory, Department of Physics, University of Cambridge, Cambridge CB3 0HE, United Kingdom
}
\author{E. T. Owen}
\thanks{These authors contributed equally to this paper.}
\affiliation{%
Institute of Photonics and Quantum Sciences, Heriot-Watt University, Edinburgh EH14 4AS, United Kingdom
}
\affiliation{%
 Cavendish Laboratory, Department of Physics, University of Cambridge, Cambridge CB3 0HE, United Kingdom
}
\author{T. Ferrus}
\affiliation{%
Hitachi Cambridge Laboratory, J. J. Thomson Avenue, CB3 0HE, Cambridge, United Kingdom
}
\author{C. H. W. Barnes}
\affiliation{%
 Cavendish Laboratory, Department of Physics, University of Cambridge, Cambridge CB3 0HE, United Kingdom
}

\date{\today}

\begin{abstract}

In this paper we present a protocol for the implementation of a positive-operator-valued measure (POVM) on massive fermionic qubits. We present methods for implementing non-dispersive qubit transport, spin rotations and spin polarizing beam-splitter operations. Our scheme attains linear optics-like control of the spatial extent of the qubits by considering groundstate electrons trapped in the minima of surface acoustic waves in semiconductor heterostructures. Furthermore, we numerically simulate a high-fidelity POVM that carries out Procrustean entanglement distillation in the framework of our scheme, using experimentally realistic potentials. Our protocol can be applied, not only to pure ensembles with particle pairs of known identical entanglement, but also to realistic ensembles of particle pairs with a distribution of entanglement entropies. This paper provides an experimentally realisable design for future quantum technologies.


\end{abstract}

\pacs{Valid PACS appear here}
\maketitle


\section{Introduction}

In quantum mechanics, the theory of measurement is far from straightforward.  Whilst there is considerable debate about the interpretations of quantum mechanics, there remain simple questions about how to formulate a mathematical description of the outcomes of recent experiments. It is often assumed that a measurement apparatus implements von Neumann's projective measurements, whereby a quantum state $\ket{\psi}$ is projected onto the eigenbasis of an observable operator $\hat{A} = \sum_i \ket{A_i} \bra{A_i}$ and the final system is measured to be in the state $\ket{A_i}$ with probability $|\langle A_i | \psi \rangle |^2$. However, in recent years, theory and experiment have shown that this definition of measurement is too restrictive.  Projective measurements fail to describe a broad range of fascinating quantum phenomena including non-demolition \cite{Braginsky80}, weak \cite{Aharanov88}, and continuous measurements \cite{Doherty99}.

A crucial component for a generalised theory of quantum measurement, is the positive-operator-valued measure (POVM).  These measures consist of a set of semi-definite non-negative operators, each associated with a particular measurement outcome, acting on the relevant Hilbert space.  By operating on a system with a POVM, followed by traditional projective measurements, it is possible to access information about a system which cannot be obtained using projective measurements alone (e.g. distinguishing between non-orthogonal states).  The uncertainty principle is maintained by allowing a finite probability that no information about the system is collected.  POVMs also have a number of applications in quantum technologies\cite{Nielsen11, IVANOVIC1987, PERES1988, Ahnert05}, contributing crucial components to entanglement distillation, quantum cryptography and quantum metrology protocols

Experimental demonstrations of POVMs have been reported in photonic systems \cite{ Kwiat01, Zhao15} but, to date, there have been no realisations of POVMs acting on particles with mass.  Whilst photons propagating in free space are non-dispersive, the wavefunction of a massive particle spreads out unless placed in a sufficiently strong confining potential. As POVMs are typically generated with quantum self-interference effects~\cite{Ahnert05},  the dispersion of massive particles is undesirable, as it reduces the fidelity of the interference.  The ability to mimic devices from quantum optics, such as the POVM, in systems where quantum information is encoded on massive particles is particularly important for the development of quantum information processing routines in solid state systems~\cite{Bocquillon12, Fletcher13, Waldie15, Kataoka16}.  For example, surface acoustic waves (SAWs) propagating on the surface of a piezoelectric semiconductor can both capture and transport electron qubits in electrostatically-defined dynamic quantum dots.  Experimentally, beam splitters~\cite{Kataoka09} and polarization readout devices \cite{Elzerman04} have been implemented in GaAs heterostructures and protocols for realising universal quantum computations have been proposed~\cite{Barnes00}.  The potential to integrate multiple components on-chip opens the possibility for developing sophisticated quantum optics-like experiments in solid-state devices.

In this paper, we present a protocol for implementing POVMs on massive electron spin-$\frac{1}{2}$ qubits.  The protocol is based on the nested polarizing Mach-Zehnder interferometer proposed by Anhert and Payne~\cite{Ahnert05} but adapted for use in a solid state setting.  We tailor a Hamiltonian to eliminate the spatial dispersion of electrons when passing through the Mach-Zehnder interferometer and the single qubit gates in order to achieve high fidelity POVMs. The spatial qubit translations are generated by SAW potentials, whilst the single qubit operations are executed with static magnetic fields. Our framework for massive particle POVMs provides a methodology for the implementation of standard optical operations on massive qubits. As an example, we demonstrate a protocol for Procrustean entanglement distillation \cite{Bennett96} of an electron spin-qubit system.

\section{\label{sec:level1} POVM Framework}

\begin{figure}
\centering
\includegraphics[scale=0.22]{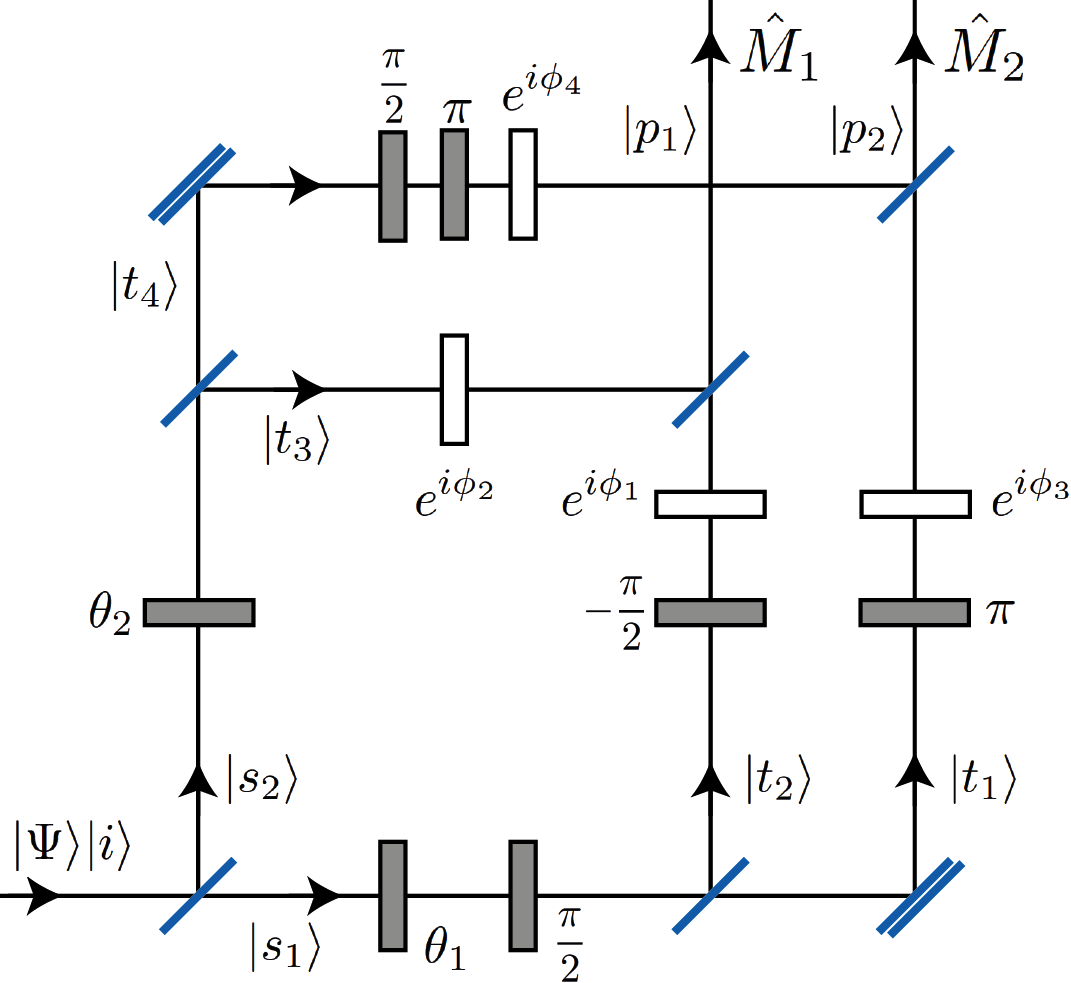}
\caption{AP POVM from reference \cite{Ahnert05}. Qubit rotations of the POVM are denoted by shaded rectangles, the phase shifts by open rectangles and the spatial degrees of freedom by the states $\ket{i}$, $\ket{s_{1,2}}$, $\ket{t_{1,2,3,4}}$ or $\ket{p_{1,2}}$. Single and double diagonal lines indicate polarizing beam-splitters and reflecting mirrors respectively.}
\label{fig:POVM}
\end{figure}

A variety of techniques have been proposed \cite{Stephen00, Ahnert05, Ahnert06} and demonstrated \cite{Wu09, Becerra13} for POVMs in optical systems.  In this paper, we use the double interferometer device proposed by Ahnert and Payne (AP)~\cite{Ahnert06} as a template from which to develop a POVM for massive particles.  Their implementation consists of two nested polarizing Mach-Zehnder interferometers  which are joined by polarizing beam splitters.  Local operations are performed on the polarization state of the photon qubit in the different arms of the interferometer using electro-optical phase modulators and wave plates as shown in Fig. \ref{fig:POVM}.  

A photon entering the system with a polarization state $\ket{\Psi} = \alpha \ket{0} + \beta \ket{1}$ leaves the interferometer in a superposition of spatial states: 
\begin{equation}
\sum_{j}{\ket{\Psi_j}\left| p_j \right\rangle} = \sum_{j}{\hat{M}_j \ket{\Psi}\left| p_j \right\rangle} ,
\end{equation}
where
\begin{align}
  \hat{M}_1 &= \cos{(\theta_1)}e^{i\phi_1}\ket{0}\!\!\bra{0} + \cos{(\theta_2)}e^{i\phi_2}\ket{1}\!\!\bra{1}  \nonumber \\
  \hat{M}_2 &= \sin{(\theta_1)}e^{i\phi_3}\ket{0}\!\!\bra{0} + \sin{(\theta_2)}e^{i\phi_4}\ket{1}\!\!\bra{1}  \nonumber
\end{align}
are the Kraus operators of the POVM.  The states $\ket{p_1}$ and $\ket{p_2}$ denote the spatially decoupled output paths such that a specific Kraus operation is performed on the polarisation state of the photon, conditioned on whether the photon exits the interferometer from output $\ket{p_1}$ or $\ket{p_2}$.  Non-diagonal Kraus operators can be created by applying unitary operations to the input and outputs of Fig. \ref{fig:POVM}. Note that generally $\hat{M}_1 \hat{M}_2 \ket{\Psi}  \neq 0$ and $\hat{M}_1 \hat{M}_1 \ket{\Psi}  \neq \hat{M}_1 \ket{\Psi}$. The operators are not necessarily orthogonal and a POVM is different from a projective operation. Whilst the  Kraus operators must satisfy:
\begin{equation}
\sum{\hat{M}^\dagger_i \hat{M}_i} = \hat{1} ,
\end{equation}
the individual Kraus operators, $ \hat{M}_i$, are not necessarily unitary.

\section{Unitary Evolution of a Massive Particle}
\label{sec:MassiveParticleOpticalAnalogues}

The interferometric scheme presented in Sec.~\ref{sec:level1} provides a template for demonstrating POVMs.  In order to map the AP POVM to a fermionic system, we will present processes which describes the individual unitary operations shown in Fig.~\ref{fig:POVM} for spin-$\frac{1}{2}$ qubits in semiconductor heterostructures.  This provides us with a toolkit allowing us to perform coherent particle propagation, spin rotations and spin-dependent particle translation on massive particles.

\begin{figure}
\centering
\includegraphics[scale=0.24]{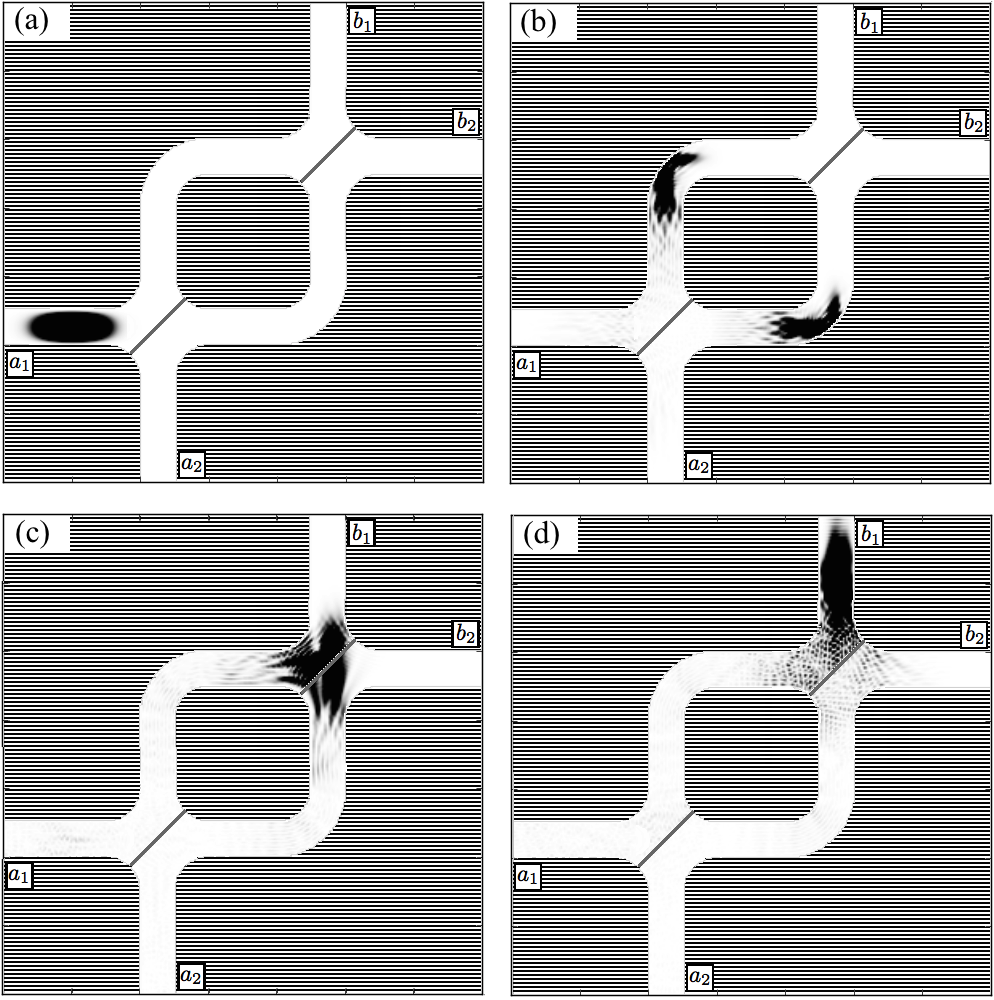}
\caption{Mach-Zehnder interferometer for massive particles at four different time steps. The potential is infinite in the striped area and zero elsewhere. The beam-splitters of the MZI are indicated with grey diagonal lines. }
\label{fig:MZIstand}
\end{figure}

The transformation of spatial propagation from photonic to fermionic states is not straightforward. Whilst a photon can pass through free space without dispersing significantly, the wavefunction of a massive particle---such as an ion or an electron---will disperse.  As most optical devices, including the polarizing Mach-Zehnder Interferometer (MZI), rely on self-interference of spatially well-defined qubit states, these systems are especially sensitive to dispersion.  Fig. \ref{fig:MZIstand} shows a staggered leapfrog \cite{Goldberg67, Askar78, Maestri00, Owen12, ArvShu16} time-evolution for the wavepacket of a massive particles passing through a MZI, where spin-dependent beam-splitters have been inserted at the junctions.  The device curvature, wavepacket shape and momentum distribution have been chosen to maximise the output probability density in the upper right port (labelled by $b_1$ in Fig. \ref{fig:MZIstand}) of the polarizing MZI. Nevertheless, over $5 \%$ of the probability density disperses to unwanted locations of the MZI and the  shape of the wavepacket is significantly distorted.  This places an upper bound of 95\% on the spin-qubit fidelity of a single polarizing MZI. Additionally, the AP POVM relies on the spatial separation of the output states and any distortion of the spatial wavepacket will inhibit optimal control. The dramatic reduction in the fidelity of the qubit operation presents a challenge for the implementation of quantum protocols, highlighting the need for a more sophisticated approach.

The optical diagram in Fig~\ref{fig:POVM} can be broken down into three separate components: the free dispersiveless propagation of the photon through the interferometer, arbitrary polarization rotation using a combination of birefringent wave plates and the spatial separation of the photon into a pair of polarized modes using polarizing beam splitters.  To replicate the AP POVM, we need to find massive particle analogues for each of these processes.

In order to realise high fidelity POVMs on massive particles, the dispersion of the states has to be eliminated. This can be obtained with Gaussian wavepackets in harmonic confining potentials. Such potentials have been used successfully in ion traps to perform coherent diabatic ion transport \cite{Bowler12, Hucul08} but equivalent potentials can be achieved in semiconductors by either electrostatically defining quantum dots using Schottky surface gates~\cite{Davies95, Tilke01, Kataoka09, Veldhorst14} or lithographically confining charges in doped regions separated by tunnel barriers \cite{Ferrus11}.

Our staggered leapfrog simulations confirm that spatial propagation can be obtained in a manner that both preserves the fidelity of the operation and keeps the shape of the wavepacket intact. There are two main ways of realising this. Firstly, the minima of the harmonic potentials can be shifted, displacing the wave packet and generating a coherent state. By imposing a diabatic shift of the ground-state potential of a stationary wavepacket, the particle can be captured when it coherently reaches the other side of the minimum of the intermediary potential. Secondly, by moving the minima of the harmonic potentials in an adiabatic manner it is possible to preserve the structure of the ground state whilst the qubit is moved between the optical component analogues.

However, our simulations show that an optimal way to adiabatically transport electrons non-dispersively is to use propagating SAW potentials. A ground-state electron (near Gaussian) trapped in the minimum of a sinusoidal SAW potential is transported coherently through the device with the propagation speed of the SAW. We suggest the use of surface Schottky gates to impose an overlying potential structure that adiabatically shifts the center of mass of the ground state in the SAW frame of reference. This effectively enables linear-optics like spatial control of the electron qubits. GPU-boosted staggered leapfrog \cite{Goldberg67, Askar78, Maestri00, Owen12} simulations allow for the parameters of the potential to be optimised for the implementation of a specific POVM.

Arbitrary polarization rotations for spin-$\frac{1}{2}$ particles can be described by time-ordered unitary operators: 
\begin{equation}
\hat{R}_{\hat{k}} = {\cal T} \exp{\big[ i \lambda(t) \sigma_{\bm{\hat{k}}} t \big]} ,
\label{Eq:Rot}
\end{equation}
where $\lambda(t)$ is some time-dependent strength parameter and $\sigma_{\bm{\hat{k}}}$ are the Pauli matrices.  Such unitary operations can be realised using a magnetic field with the Hamiltonian $\hat{H}_{rot} = - \bm{\mu} \cdot \bm{B}(t)$, where $\bm{\mu}$ is the magnetic dipole moment of the particle and the magnetic field $\bm{B}(t)$ is uniform over the particle wavepacket \cite{Furuta04, McNeil10}. Spin-rotations of SAW qubits have been studied in previous works \cite{Barnes00}. Charged qubits moving in a magnetic field will naturally experience a Lorentz force. However, for SAW carried electrons in semiconductor heterostructures, this force is counteracted greatly by the device confinement. Electromagnetic corrections can also be applied as suggested in \cite{Barnes00}. Other techniques for spin rotations include using a DC magnetic field to lift the spin degeneracy and applying an oscillating perpendicular magnetic field set in resonance between the two spin states \cite{Awschalom13}. Yet another technique uses electron spin resonance (ESR), where a pulse of microwaves becomes resonant with the upper and lower Zeeman-split spin states \cite{Kawakami16}.

Although solid-state physics present several possibilities to select the spin of an electron (Pauli blockade \cite{Prati11} or spin filtering \cite{Pla12}), implementing a spin-splitter is difficult in practice, owing to the generally small dimensions of devices and the intrinsic nature of the spin. However, several structures, materials or techniques can be used to channel dedicated spin orientations. 

Antidots \cite{Zozoulenko04} or quantum spin hall systems \cite{Kato04} are commonly used to create spin-polarised channels at the edges of structures with a minimum number of gates and simplified geometry. These have been realised in graphene \cite{Tada12} but also in semiconductors. More generally, it is possible to utilise materials with strong spin-orbit interaction to generate spin currents out of charge current.  Another approach is to scatter the wave packet off of a narrow magnetic semiconductor barrier, such as EuO~\cite{Santos07}, which will act as a spin filter only transmitting a specific electron spin polarisation.  Furthermore, new types of materials, like topological insulators, possess intrinsic properties that allow locking spin states to specific transport directions \cite{Konig07}.


Finally, there exist a number of schemes for the projective measurement of fermion spin \cite{Kane98, Shnirman98, Makhlin99, Gardelis99, Barnes00, Elzerman04, Weber14}. These schemes implement spin-dependent translations of the qubits followed by a single particle charge readout. Technologies for projective spin measurements are based on magnetic readout (utilising the spin-valve effect), double occupation readout (utilising spin-dependent tunneling) or Stern-Gerlach readout.

\section{ Massive Particle POVM}

With this massive particle toolkit, we provide a proof-of-principle simulation of a fermionic POVM.  Whilst our protocol can be used to implement any POVM on the massive spin-$\frac{1}{2}$ particle, we use the implementation of an entanglement distilling POVM as a guiding example in this section.

\subsection{Procrustean Entanglement Distillation}

One use of POVMs is found in the implementation of Bennett's Procrustean entanglement distillation.\cite{Bennett96} This protocol allows a subset of pure state qubit pairs to be discarded from a weakly entangled ensemble, such that the remaining particle pairs are more entangled. Significantly, Bennett's method can be { \it local} and {\it non-iterative} as the entanglement distillation is achieved through the application of a single POVM on only one of the particles.

For the arbitrarily entangled state,
\begin{equation} \label{eq:in}
\ket{\Psi_{\rm{A,B}}} =
\alpha \ket{0_{\rm{A}}}\!\ket{0_{\rm{B}}} +
\beta \ket{1_{\rm{A}}}\!\ket{1_{\rm{B}}},
\end{equation}
shared between say Alice and Bob, Procrustean entanglement distillation can be achieved by applying a POVM to just Alice's particle, creating the maximally entangled Bell state:
\begin{equation}
\ket{\Psi_{\rm{A,B}}} =
\frac{1}{\sqrt{2}} \Big( \ket{0_{\rm{A}}}\!\ket{0_{\rm{B}}} \pm
 \ket{1_{\rm{A}}}\!\ket{1_{\rm{B}}} \Big) .
\end{equation}
with probability $P_{dist} = 2(1-\max(|\alpha|^2, |\beta|^2))$.

\subsection{ POVM Parameters for Distillation}
\label{subsec:POVMParams}

The parameters for the massive particle POVM can be adjusted to carry out the Procrustean entanglement distillation protocol described above. We introduce two new parameters $\varphi$ and $\gamma$ which, for a known initial state of the form of Eq. \ref{eq:in}, are set such that $\alpha \equiv \cos{(\varphi)}$ and $\beta \equiv \exp({i\gamma})\sin{(\varphi)}$. The POVM parameters are then set according to Table \ref{tab:Param}.   Alice inserts a detector at the $\ket{p_2}$ output and passes her particle through the POVM. The wavefunction output at $\ket{p_1}$, is then acted on by the operator $\hat{M}_{1}^A = \tan{(\varphi)}\ket{0}\!\bra{0} + \ket{1}\!\bra{1}$ if $ l\pi - \pi/4 \leq \varphi \leq l\pi + \pi / 4$ (for integer $l$), and $\hat{M}_{1}^A = \ket{0}\!\bra{0} + \cot{(\varphi)}\ket{1}\!\bra{1}$ otherwise. The two-particle state is output as $\ket{\Psi_1}=\frac{1}{\sqrt{2}} (\ket{1_{\rm{A}}}\!\ket{1_{\rm{B}}} + \ket{0_{\rm{A}}}\!\ket{0_{\rm{B}}})$ with probability $P_1=1-|\cos{(2\varphi)}|= 2(1-\max(|\alpha|^2, |\beta|^2))$. The choice of these parameters allows Alice to locally distill the entanglement she shares with Bob, by passing her particle ensemble through the device in Fig. \ref{fig:POVM}. The successful creation of a Bell state at the  $p_1$-output can be heralded by the lack of detection of a particle at the $p_2$-output.

\begin{table}[h!]
\caption{\label{tab:Param}%
POVM parameters for the implementation of entanglement distillation of the state in Eq. \ref{eq:in}.
}
\begin{ruledtabular}
\begin{tabular}{cccc}
 & $\phi_{1}$ & $0$  &  \\
 & $\phi_{2}$ & $0$   & \\
 & $\phi_{3}$ & $-\gamma$   & \\
 & $\phi_{4}$ & $-\gamma$   & \\
 & $\theta_1$ & $\Re{[\arccos{(\tan{(\varphi)})}]}$  &  \\
 & $\theta_2$ & $\Re{[\arccos{(\cot{(\varphi)})}]}$  & 
\end{tabular}
\end{ruledtabular}
\end{table}

\subsection{ Simulating the Massive Wavepacket Evolution}

Using the single-qubit operations of Section \ref{sec:MassiveParticleOpticalAnalogues}, the implementation of our POVM for spin-$\frac{1}{2}$ particles in a SAW system can be simulated. By setting the POVM parameters in accordance with Sec.~\ref{subsec:POVMParams}, we implement Procrustean entanglement distillation on a massive wavepacket. 


In this section, we demonstrate the implementation of a POVM that distills the entanglement by operating on a single particle from a joint initial state of the form $\ket{\Psi_{A,B}}=\cos{(60\degree)}\ket{\downarrow_A}\ket{\downarrow_B}+i \sin{(60\degree)}\ket{\uparrow_A}\ket{\uparrow_B}$. The spatial degree of freedom is labelled by $\ket{i}$, $\ket{s_{1,2}}$, $\ket{t_{1,2,3,4}}$ and $\ket{p_{1,2}}$, as in Fig. \ref{fig:POVM}.


\onecolumngrid

\begin{figure}[h]
\centering
\includegraphics[scale=0.62]{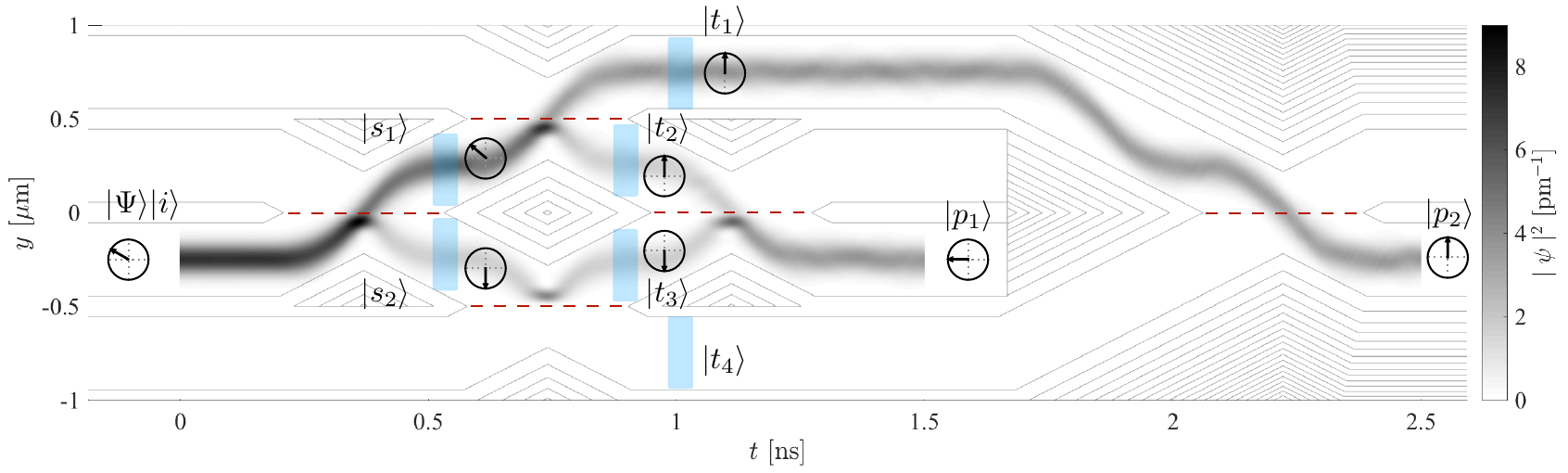}
\caption{Simulation of a massive wavepacket travelling through a POVM device. The electrostatic potential of the proposed semiconductor device is represented by grey contour lines. Dashed lines indicate the position of the spin beam-splitters. After each beam-splitter, a magnetic field, represented by the shaded areas, is applied for spin rotations according to Fig. \ref{fig:POVM}. The arrow in the projected Bloch spheres indicate the wavefunction's spin orientation in their respective regions. In the simulation presented, the electron wave function is split with equal probability between the $\ket{p_1}$ and $\ket{p_2}$ outputs.}
\label{fig:POVMSIM}
\end{figure}

\twocolumngrid
An overlying potential is necessary in order to achieve the confinement necessary for the double interferometer. It can be implemented with Schottky gates, as described above, or by etching the semiconductor material. The contour lines in Fig. \ref{fig:POVMSIM} show such an electrostatic potential. The sinusoidal SAW potential is not included in the figure.

In Fig. \ref{fig:POVMSIM} the two-dimensional electron wavefunction is traced out in the $x$-dimension, showing the probability distribution in the $y$-dimension as a function of time, $t$. Because of the strong confinement of the SAW potential, the particle distribution and movement in the $x$-direction is minimal. Hence, its $x$-position can be accurately estimated by $x = v \cdot t$, where $v$ is the speed of sound in the material.

The electron initially exists in the ground state of the SAW minimum, in the spatial state $\ket{i}$. The direction of motion is changed, and it is incident on the first polarizing beam-splitter. Here the electron is split into its spin components in a superposition of the spatial states $\ket{s_1}$ and  $\ket{s_2}$. Two magnetic fields are applied to the respective components indicated by the shaded areas in Fig. \ref{fig:POVMSIM}. $\ket{s_1}$ and  $\ket{s_2}$ are then incident on two beam-splitters forming a new superposition of the states $\ket{t_1}$, $\ket{t_2}$, $\ket{t_3}$ and $\ket{t_4}$ ($\ket{t_4}$ is not occupied for this specific POVM). Again, magnetic fields (shaded areas) are applied to implement local phase shifts and spin-rotations on the individual spatial components of the electron. Following these magnetic fields, the spatial components $\ket{t_2}$ and $\ket{t_3}$ are interfered on a beam-splitter, forming an output component $\ket{p_1}$. Similarly, $\ket{t_1}$ and $\ket{t_4}$ are interfered to form $\ket{p_2}$. 

Hence, Fig. \ref{fig:POVMSIM} shows how an input wavefunction $\ket{\psi_A}\ket{i}$ is transformed into a spatial superposition given by $\hat{M}_{1}\ket{\psi_A}\ket{p_1}+\hat{M}_{2}\ket{\psi_A}\ket{p_2}$. In a 2D structure, $\ket{p_1}$ has to be trapped such that $\ket{t_1}$ and $\ket{t_4}$ can evolve around it. However, recent successes in creating rolled-up semiconductor nanotubes \cite{Prinz00, Schmidt01, Brick17} and layered quantum well structures \cite{Laikhtman09,Sivalertporn12,Cohen16} would allow output arms to continue to evolve through space, by enabling “periodic” boundary conditions, and finite 3D movement respectively.

By utilising the stability of a wavepacket carried by a SAW, and by optimising the device parameters, our simulations are able to demonstrate experimentally achievable high fidelity POVMs. Moreover, whilst this subsection has demonstrated a specific implementation, the extension to a general POVM with more than two Kraus operators is straightforward \cite{Ahnert06}.  Nested polarizing Mach-Zehnder interferometers can be connected together by inserting the output states at $\ket{p_1}$ and $\ket{p_2}$ into subsequent interferometers in order to generate a POVM with any combination of Kraus operators.

\subsection{ Distillation of Realistic Distributions of Entangled Particle Pairs}

We have shown how a POVM can be implemented on massive spin-$\frac{1}{2}$ qubits.  The Procrustean distillation protocol assumes that the initial pure state is known.  Experimentally, it is likely that processes which produce entangled massive states produce ensembles of particle pairs with a distribution of entanglement strengths.   Whilst there exist theoretical methods for the entanglement distillation and purification of mixed states~\cite{Bennett96-3, Murao98, Pan01}, these methods are iterative and require two-qubit operations.  Owing to the experimental difficulties in the application of such operations, it is valuable to investigate the effect of the non-iterative single-qubit protocol on realistic particle pair ensembles. 

By selecting a subset of the particles from the ensemble, one can optimise the POVM configuration to maximise the entropy of entanglement of the pairs in the final ensemble. The subset of particles used in the optimisation is consumed. However, the remaining ensemble can pass through the optimised POVM, in order to generate a reduced ensemble of higher pairwise entanglement.

\begin{figure}
\centering
\includegraphics[scale=0.15]{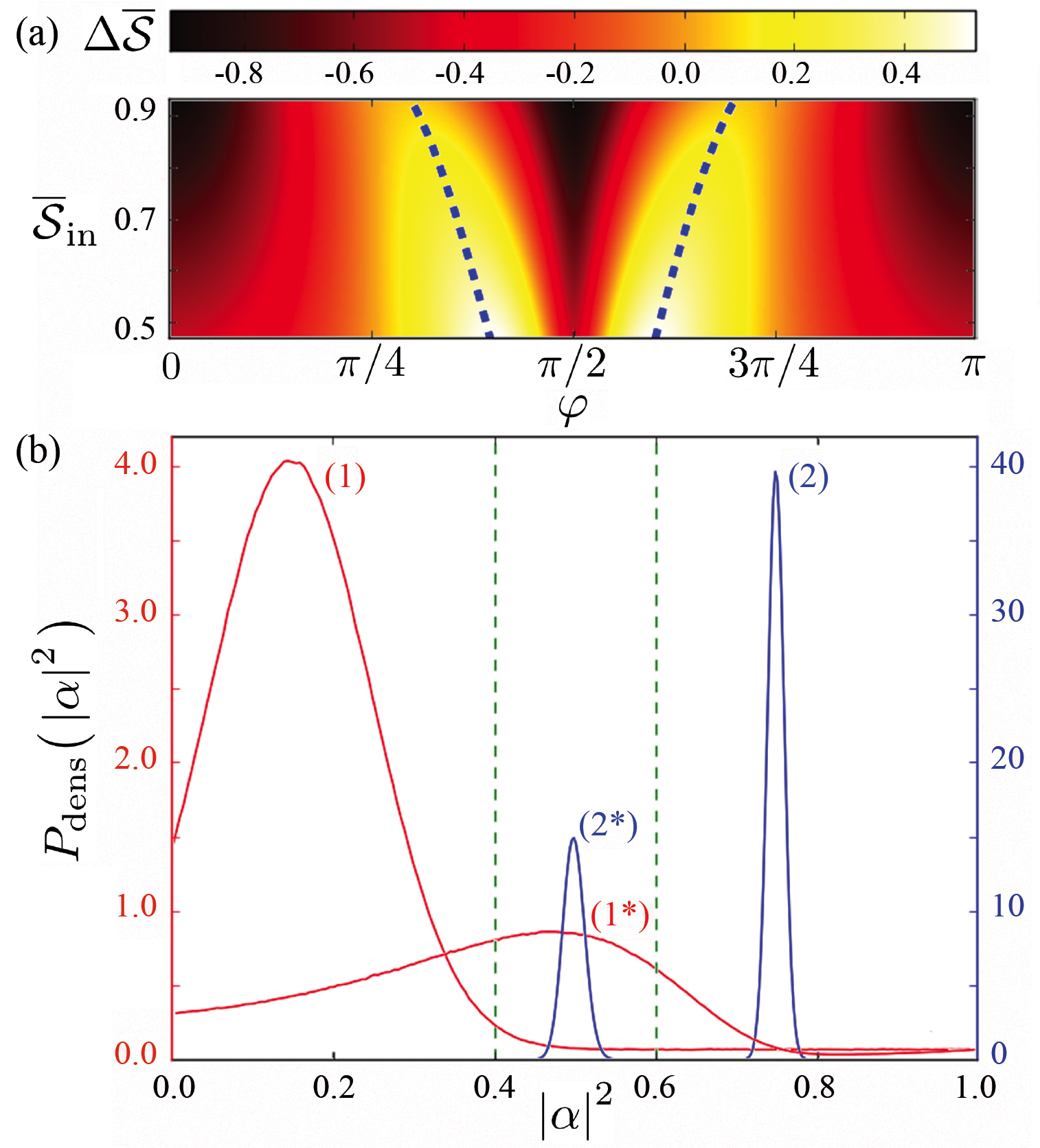}
\caption{(color online) (a) Difference between the initial and the final ensemble mean entropy of entanglement (contour from color-bar). The horizontal axis shows the POVM parameter, $\varphi$, and the vertical axis shows the initial mean value of the entropy of entanglement. (b) Probability density as a function of $|\alpha|^2$, of two example input distributions, (1) and (2), and their corresponding non-normalized $\ket{p_1}$ output distributions, (1*) and (2*).}
\label{fig:DiffProbs}
\end{figure}

In Fig. \ref{fig:DiffProbs} we show the difference in the von Neumann entanglement entropy distribution for particle pair ensembles before and after the distillation protocol.  Fig. \ref{fig:DiffProbs}(a) shows the change in the mean entanglement entropy, $\Delta \overline{\mathcal{S}}$, as a function of initial mean entropy, $\overline{\mathcal{S}}_{\rm{in}}$, and POVM angle, $\varphi$, as previously related to $\theta_1$ and $\theta_2$. We have assumed that the value of $|\alpha|^2$ (the probability of state $\ket{0}$) in the initial particle pairs has a Gaussian profile of width $\sigma=0.01$. The dotted lines show the loci of the optimal POVM angles for ensembles of identical pairs. A lower  initial mean entanglement allows for the possibility of a higher increase of mean entanglement. The simulations were carried out using Monte Carlo theory with an ensemble size of $10^5$ particles in each distribution.  For well-behaved distributions, the proposed setup will efficiently produce a final ensemble of increased average entanglement. This is true even for wide distributions such as curve (1) in Fig. \ref{fig:DiffProbs}(b).

\section{ Concluding Remarks }

We have developed a methodology for the implementation of massive spin-$\frac{1}{2}$ qubit POVMs. The POVM builds on the framework of the AP double interferometer POVM \cite{Ahnert05}.  We have proposed a toolkit for translating the optical components from the AP POVM into processes which are suitable for electrons in surface acoustic wave systems.  The use of ground state wavefunctions of SAW minima allows us to virtually eliminate the dispersion of the particle wavepackets, providing the means to replicate the optical POVM with a massive particle analogue.  Owing to the difficulty in controlling photon-photon interactions, linear-optics-like processing of massive (more easily interacting) particles will be valuable for quantum computational aspirations or quantum cryptography with hybrid systems.

We demonstrated the effectiveness of the proposed scheme by simulating the evolution of a spin-$\frac{1}{2}$ POVM that performs Procrustean entanglement distillation on a pair of entangled massive qubits. Using a Hamiltonian tailored by GPU-boosted parameter sweeps, our simulation showed a POVM fidelity of $>99.5 \%$. However, this is not an upper bound and additional parameter optimisation can lead to even higher fidelities. Furthermore, our Monte-Carlo based numerical investigation shows how the protocol can increase the average entropy of entanglement of particle pair ensembles with distributions of initial entanglement entropies.

\section{Acknowledgements}

The authors would like to express their gratitude towards Aleksey Andreev for helpful discussions. This work was supported by the EPSRC, Cambridge Laboratory of Hitachi Limited via Project for Developing Innovation Systems of the MEXT in Japan, Tornspiran's Trust Fund, Lars Hierta's Memorial Foundation, L\"angmanska kulturfonden's Trust Fund, Sixten Gemzeus' Fund and	Anna Whitlock's Fund. Furthermore, it has received funding from the European Union's Horizon 2020 research and innovation programme under the Marie Sk\l{}odowska-Curie grant agreement No 642688.

\bibliography{EntDist}

\end{document}